\documentclass[preprint,preprintnumbers,prl]{revtex4}
\usepackage{graphicx}
\usepackage{dcolumn}
\usepackage{bm}

\begin{document}


\title{Anomalous Hall Effect and Magnetic Monopoles in Momentum-Space}

\author
{Zhong Fang$^{1,2}$\footnote{To whom correspondence should be
addressed. Email: z.fang@aist.go.jp}, 
Naoto Nagaosa$^{1,3,4}$, Kei S. Takahashi$^{5}$,
Atsushi Asamitsu$^{1,6}$,
Roland Mathieu$^{1}$, Takeshi Ogasawara$^{3}$, Hiroyuki Yamada$^{3}$, 
Masashi Kawasaki$^{3,7}$, Yoshinori Tokura$^{1,3,4}$, 
Kiyoyuki Terakura$^{8}$\\  \  \\ }

\affiliation{
$^1$Spin Superstructure Project (SSS), ERATO, 
Japan Science and Technology Corporation (JST),
AIST Tsukuba Central 4, Tsukuba 305-8562, Japan;\\
$^{2}$Institute of Physics, Chinese Academy of Science,
Beijing 100080, China;\\
$^3$Correlated Electron Research Center (CERC), AIST Tsukuba
Central 4, Tsukuba 305-8562, Japan;\\
$^4$Department of Applied Physics, University of Tokyo, 7-3-1, Hongo,
Bunkyo-ku, Tokyo 113-8656, Japan;\\
$^5$DPMC, University of Geneva, 24, quai Ernest-Ansermet, 1211 Geneva
4, Switzerland;\\
$^6$Cryogenic Center, University of Tokyo, 2-11-16 Bunkyo-ku, 
Tokyo 113-0032, Japan;\\
$^7$Institute for Materials Research, Tohoku University, 
Sendai 980-8577, Japan \\
$^8$Research Institute for Computational Sciences (RICS),
AIST Tsukuba Central 2, Tsukuba 305-8568, Japan}

\maketitle

{\bf{Efforts to find the magnetic monopole in real space have been made in
cosmic rays and in accelerators, but up to now there is no firm
evidence for its existence due to the very heavy mass $\sim
10^{16}$GeV.  However, we show that the magnetic monopole can appear
in the crystal-momentum space of solids in the accessible low energy
region ($\sim0.1-1$eV) in the context of the anomalous Hall effect.
We report experimental results together with first-principles
calculations on the ferromagnetic crystal SrRuO$_3$ that provide
evidence for the magnetic monopole in the crystal-momentum space.}}

In 1931, Dirac \cite{dirac} postulated the existence of magnetic
monopole (MM) searching for the symmetry between the electric and
magnetic field in the laws of electromagnetism.  The singularity of the
vector potential is needed for this Dirac magnetic monopole.
Theoretically, the MM was found \cite{monopole1,monopole2} as the
soliton solution to the equation of the non-Abelian gauge theory for
the grand unification.  However its energy is estimated to be
extremely large $\sim 10^{16}$GeV, which makes its experimental
observation difficult.  In contrast to this MM in real space, one can
consider its dual space, namely the crystal momentum (${\bf k}$-) space
of solids, and the Berry phase connection \cite{berry} of Bloch
wavefunctions. This magnetic monopole in momentum space is closely
related to the physical phenomenon, i.e., the anomalous Hall effect
(AHE) in ferromagnetic metals.

The AHE is a phenomenon where the
transverse resistivity $\rho_{xy}$ in ferromagnets contains the
contribution due to the magnetization $M$ in addition to the usual
Hall effect.  The conventional expression for $\rho_{xy}$ is
\begin{equation}
\rho_{xy} = R_0 B + 4 \pi R_s M
\end{equation}
where $B$ is the magnetic field, $R_0$ is the usual Hall coefficient,
and $R_s$ is the anomalous Hall coefficient.  This expression
implicitly assumes that the additional contribution is proportional to
the magnetization, and is used as an experimental tool to measure the
magnetization as a function of temperature. Especially it is
extensively used in the studies of ferromagnetic semiconductors with
dilute magnetic impurities, which are the most promising materials for
the spintronics \cite{exp.}.  However, the mechanism of AHE has been
controversial for long \cite{KL,smit,kohn,luttinger58,berger}.  The
key issue is whether the effect is intrinsic or extrinsic and how to treat
the impurity and/or phonon scatterings.  Karplus and Luttinger
\cite{KL} was the first to propose the intrinsic mechanism of AHE,
where the matrix elements of the current operators are essential.
Other theories \cite{smit,berger} attribute the AHE to the impurity
scattering modified by the spin-orbit interaction, namely the skew
scattering \cite{smit} and/or the side-jump mechanism \cite{berger}.
These extrinsic mechanisms are rather complicated and depends on the
details of the impurities as well as the band structure. Nevertheless,
all those conventional theories \cite{KL,smit,kohn,luttinger58,berger}
for AHE derives Eq.1, as they are based on the perturbative expansion
in the spin-orbit coupling (SOC) $\lambda$ and the magnetization $M$,
i.e., $R_s \propto \lambda $.

Recently the geometrical meaning of the AHE of the intrinsic origin
\cite{KL} has been recognized by several authors
\cite{onoda,Niu,taguchi,mac}. The transverse conductivity
$\sigma_{xy}$ can be written as the integral of the Berry phase
curvature (gauge field) over the occupied electronic states in 
crystal momentum space (Eq.5 below).  Magnetic monopole 
corresponds to the source or
sink of the gauge field/curvature defined by this Berry
phase connection.  Therefore the AHE can be the direct fingerprint of
the MM in crystal momentum space. Note here that the presence of the
time-reversal symmetry results in $\sigma_{xy} = \rho_{xy} = 0$ 
in the d.c. limit from the very generic argument, and the group theoretical
condition for the nonzero $\sigma_{xy}$ is equivalent to that of
finite ferromagnetic moment (see Supporting Online Material (SOM)).  
Therefore we need the ferromagnets to study $\sigma_{xy}$ even 
though the Berry phase connection is more universal and exists 
even in nonmagnetic materials.

In this paper, we show by detailed first principles band calculation
combined with the transport, optical, and magnetic measurements, that
the observed unconventional behavior of the anomalous Hall effect and
Kerr rotation in metallic ferromagnet SrRuO$_3$ is of intrinsic origin
and is determined by the MM in ${\bf k}$-space.  The conventional
expression Eq.1 is invalidated by the experimental data presented
below, showing the non-monotonous temperature dependence including
even a sign change.

SrRuO$_3$ with the perovskite structure is an itinerant (metallic)
ferromagnet. There are 4 $t_{2g}$ electrons with the low spin
configurations. The $4d$ orbitals of SrRuO$_3$ are relatively extended
and the bandwidth is large compared with the Coulombic interaction.
The relativistic SOC is also large in $4d$ electrons because of the
heavy mass (of the order of 0.3eV for Ru atom).  High-quality single
crystal is available with the residual resistivity of the order of $10
\mu \Omega$cm. These aspects make this system an ideal candidate to
observe the AHE due to the ${\bf k}$-space gauge field. Stoichiometric
SrRuO$_3$ (bulk and thin film) and Ca-doped
(Sr$_{0.8}$Ca$_{0.2}$RuO$_3$ thin film) single crystals were prepared
(see SOM for the details of sample preparation).  The magnetization
curve (Fig.1A) of SrRuO$_3$ film is quite similar to that of bulk
single crystal, except the Curie Temperature $T_c$ ($\sim$150K) is
slightly lower than that of bulk ($\sim$160K), due to the strain
effects. On the other hand, the isovalent Ca-doping suppresses the
$T_c$ and magnetization $M$ dramatically.  All of the samples were
used for the transport measurement in the d.c. limit. As seen from
Fig.1C, the $\rho_{xy}$ changes non-monotonously with temperature
including even a sign change. Such behavior is far beyond the
expectation based on the conventional expression Eq.1. In addition to
the transport measurement, the frequency ($\omega$) dependent
conductivities (Fig. 2) were measured for SrRuO$_3$ film by optical
method (see SOM). Except the strong structures around 3.0eV, which are
mostly due to the charge transfer from O-$2p$ to Ru-$4d$, sharp
structures are also observed for both the real and the imaginary part
of $\sigma_{xy}(\omega)$ below 0.5eV. Those low energy sharp
structures cannot be fitted by extended Drude analysis. The lower the
energy is, the stronger the deviations from fitting are.  We will
show, by combination with the first-principles calculations, that
those unconventional behavior actually originate from the singular
behavior of MM in the momentum space.

Now we turn to some details of the theoretical analysis.
Berry phase is the quantal phase acquired by the wavefunction 
associated with the adiabatic change of the Hamiltonian \cite{berry,book}.
Let $ | n ({\bf \alpha})>$ be the $n$-th eigenstate of the Hamiltonian 
$H({\bf \alpha})$ with ${\bf \alpha}= (\alpha_1, ....,\alpha_m)$ 
being the set of parameters, the Berry's connection is the overlap of the 
two wavefucntions infinitesimally separated in ${\bf \alpha}$-space, i.e.,
\begin{equation}
< n ({\bf \alpha}  )| 
n ({\bf \alpha} + \Delta {\bf \alpha} )>
=  1 +   \Delta {\bf \alpha} \cdot  
 < n ({\bf \alpha})| \nabla_{\bf \alpha} 
| n ({\bf \alpha})>
= \exp \biggl[ - i  \Delta {\bf \alpha} \cdot  {\bf a}_n ({\bf \alpha})
\biggr].
\end{equation}
where the vector potential ${\bf a}_n ({\bf \alpha})$ is defined by 
$ {\bf a}_n ({\bf \alpha}) = i < n ({\bf \alpha})| \nabla_{\bf \alpha}
| n ({\bf \alpha})>$. Although the concept of Berry phase has broad 
applications in physics \cite{book}, its relevance to the 
band structure in solids has been recognized only recently and in limited 
situations such as the quantum Hall effect under strong magnetic field 
\cite{TKNN} and the calculation of electronic polarization 
in ferroelectrics \cite{Martin,vanderbilt}. 
In this case the parameter ${\bf \alpha}$ is the crystal momentum  
${\bf k}$. For the Bloch wavefunction 
$\psi_{n {\vec k}}(\vec r) = 
e^{i{\vec k} \cdot{\vec r}}u_{n {\vec k}}(\vec r)$ 
with $n$ denoting the band index and $u_{n {\vec k}}$ being the periodic 
part, the vector potential for the Berry phase  $a_{n \mu} ({\vec k})$ is 
\begin{equation}
a_{n \mu }({\vec k}) = 
i < u_{n {\vec k}} | { {\partial } \over { \partial k_\mu} } u_{n {\vec k}}>.
\end{equation}
With this vector potential, the gauge covariant position operator
$x_\mu$ for the wavepacket made out of the band $n$ is 
given by $x_\mu = i \partial_{ k_\mu}  - a_{n \mu }({\bf k}) $.
Therefore the commutation relation of $x_\mu$'s includes
the gauge field $F_{\mu \nu} = \partial_{k_\mu}  a_{n \nu}
- \partial_{k_\nu}  a_{n \mu}$ as
\begin{equation} 
[ x_\mu, x_\nu ] = -i F_{\mu \nu}.
\end{equation} 
which leads to the additional (anomalous) velocity 
$ - i[ x_\mu, V(x) ] = - F_{\mu \nu} \partial V(x)/ \partial x_\nu $ 
being transverse to the electric field 
$ E_\nu = -  \partial V(x)/ \partial x_\nu $. 
Therefore the transverse conductivity $\sigma_{xy}$ is
given by sum of this anomalous velocity over the occupied states as
\cite{TKNN}
\begin{equation}
\sigma_{xy} = \sum_{n, {\bf k}} n_F( \varepsilon_{n}({\bf k}))
b_z({\bf k}).
\end{equation}
where $b_z({\bf k}) = F_{xy} ({\bf k})$ and $n_F(\varepsilon) =
1/(e^{\beta( \varepsilon - \mu)} +1)$ ( $\beta$: inverse temperature,
$\mu$: chemical potential ) is the Fermi distribution function. Hence
the behavior of gauge field $b_z({\bf k})$ in $k$-space \cite{real}
determines that of $\sigma_{xy}$.  One might imagine that it is a
slowly varying function of ${\bf k}$, but it is not the case. Fig. 3B
is the calculated result for $b_z({\bf k})$ in the real system
SrRuO$_3$.  It has a very sharp peak near $\Gamma$-point and also
ridges along the diagonals.  The origin for this sharp structure is
the (near) degeneracy and/or the band crossing, which act as MM.
Consider the general case where two band Hamiltonian matrix $H(\bf k)$
at $\bf k$ can be written as $H({\bf k}) = \sum_{\mu =0, 1,2,3}
f_\mu(\bf k) \sigma_\mu $ where $\sigma_{1,2,3}$ are the Pauli
matrices and $\sigma_0 $ is the unit matrix.  We can consider the
mapping from $\bf k$ to the vector ${\bf f}({\bf k}) = ( f_1({\bf k}),
f_2({\bf k}), f_3({\bf k}))$, and the contribution to $\sigma_{xy}$
from the neighborhood of this degeneracy region is given by the solid
angle $d \Omega_{\bf f}$ for the infinitesimal $d k_x d k_y$
integrated over ${\bf k}$.  Therefore the gauge flux near the MM,
namely the degeneracy point ${\bf f} = {\bf 0}$ is singularly enhanced
(as shown in Fig. 3). (See SOM for more details).

We studied the behavior of $\sigma_{xy}$ by the first-principles
calculations (see SOM for methods).  The calculated density of states
(DOS) is not so different between the cases with and without SOC
(Fig.4A), while the $\sigma_{xy}$ should be very sensitive to the
Bloch wavefunctions and depends on the Fermi level position and the
spin-splitting (magnetization) significantly, as predicted by above
discussions. The behavior of $\sigma_{xy}$ as a function of the
Fermi-level position was obtained by using a small broadening
parameter for the lifetime $\delta$ (=70meV) (Fig.4B). By shifting the
Fermi level, not only the absolute value but also the sign of
$\sigma_{xy}$ is found to change.  The sharp and spiky structures are
just the natural results of singular behavior of MM (Fig.3).  For the
case without any shift of Fermi level, we obtain the value of
$\sigma_{xy}=-60$ $\Omega^{-1}$cm$^{-1}$, which has the same sign as
and is comparable with the experimental value
($\sim-100\Omega^{-1}$cm$^{-1}$).  Such a spiky behavior should be
also reflected in the $\omega$ dependent $\sigma_{xy}$, especially for
the low energy range with longer lifetime, while be suppressed at
higher activation energy with shorter lifetime. As shown in Fig.2, for
the $\omega$ dependence of optical conductivity, the high energy part
($>$0.5eV), which is dominated by the $p-d$ charge transition peak, is
usual, and can be well reproduced by our calculations. While the
observed peak structure of $\sigma_{xy}(\omega)$ below 0.5eV is the
clear demonstration of the predicted spiky behavior. The spectra below
0.2eV is not measured due to the technical difficulty, but it is clear
that even sharper structure should be there because the d.c. limit
Re$(\sigma_{xy})\approx-100\Omega^{-1}$cm$^{-1}$ has the opposite sign
(the Im$(\sigma_{xy})$ at d.c. limit should go back to zero). Such a
low energy behavior is well represented by our calculations, providing
further evidence for the existence of magnetic monopoles.

Now it is straightforward to understand the results of transport
measurement for $\sigma_{xy}$. Here we attribute the temperature ($T$)
dependence of $\sigma_{xy}$ to that of the magnetization $M(T)$.  As
the results of $\vec k$-space integration over occupied states, the
calculated $\sigma_{xy}$ is non-monotonous as a function of
magnetization (Fig.1D).  With the reduction of spin-splitting, the
calculated $\sigma_{xy}$, after the initial increase, decreases
sharply, then increases and changes sign becoming positive, and
finally goes down again, capturing the basic features of the
experimental results. Even more surprisingly, by converting the
measured $\rho_{xy}$ versus $T$ curves shown in Fig.1C into the
$\sigma_{xy}$ versus $M$ curves shown in Fig.1D, now they all follow
the same trend, and match with our calculations as long as the
experimental data is available. Those curves are measured for
different samples (with different saturation moments), they all follow
the same rule qualitatively, and could be simply explained by the
reduction of magnetization~\cite{Ca-more} (see Fig.1A).  Note,
however, that the comparison between the experiments and calculations
should be semi-quantitative, because the results are sensitive to the
lattice structures.  The calculated $\sigma_{xy}$ for the fictitious
Cubic structure shows a strong deviation from that obtained for
orthorhombic structure, and it changes the sign to be positive at low
temperature (large $M$).  Therefore more accurate information on the
structure is needed to obtain the quantitative result. However, such a
sensitivity does not affect our main results, i.e, the non-monotonous
behavior of $\sigma_{xy}$. Even the calculations for Cubic structure
show such a behavior, and may be used as a guide of possible
deviation.

The results and analysis presented here should stimulate and urge the
reconsideration of the electronic states in magnetic materials from
the very fundamental viewpoint.  For example, the magnetic monopole is 
accompanied by the singularity of the vector potential, i.e., 
Dirac string \cite{dirac}. As shown by Wu-Yang
\cite{wuyang} this means that more than two overlapping 
regions have to be introduced,  in each of which the gauge of the 
wavefunction is defined smoothly. 
This means that one cannot define the phase of the Bloch
wavefunctions in a single gauge choice when the magnetic monopole 
is present in the crystal momentum space. 
This leads to some nontrivial consequences such as the 
vortex in the superconducting order parameter as a function 
of ${\bf k}$ \cite{super}, and many others are left for 
future studies.

\newpage
\begin{figure}
\includegraphics[scale=0.7]{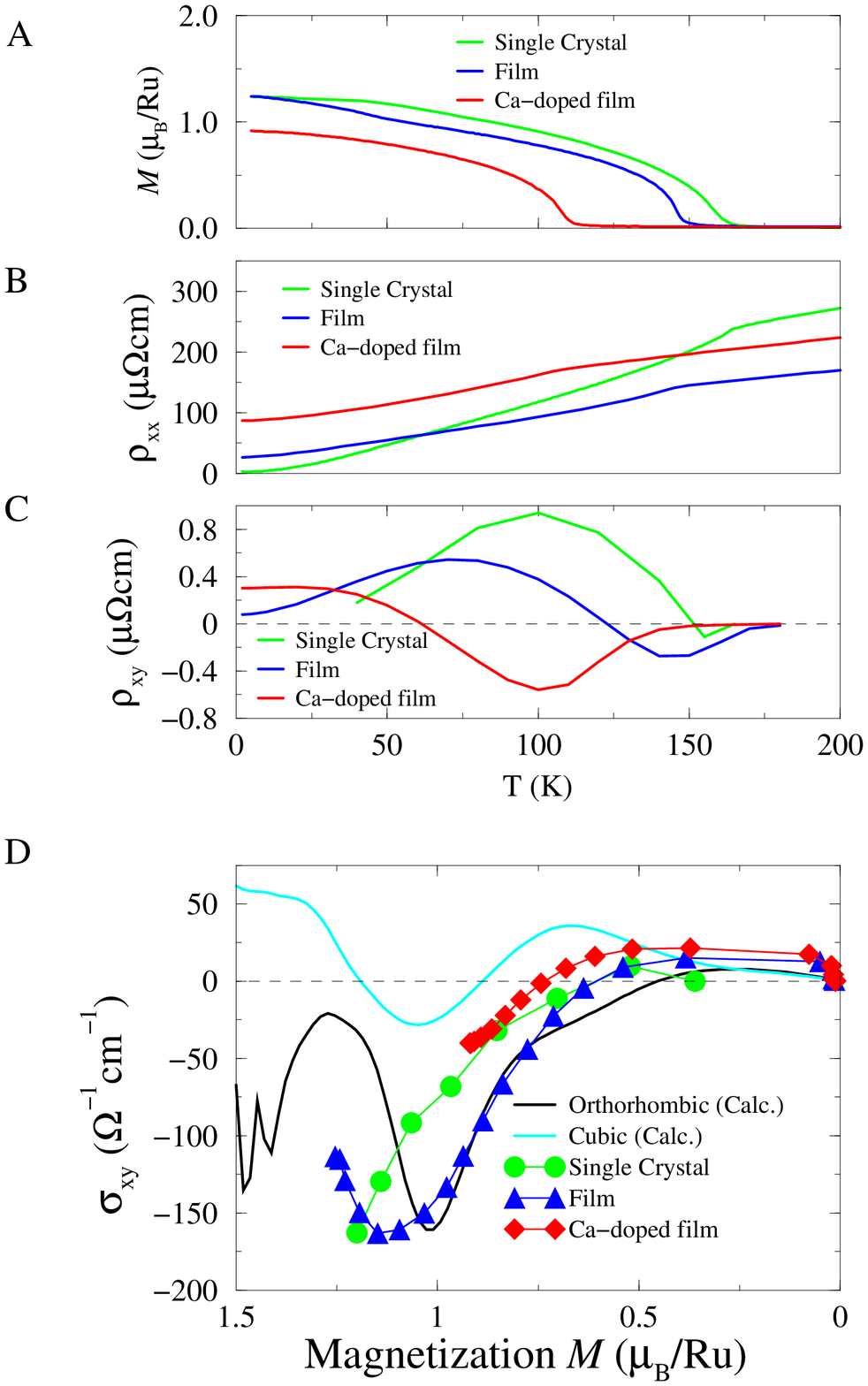}
\caption{Measured temperature dependence of the (A)
magnetization $M$, (B) longitudinal resistivity $\rho_{xx}$, and (C)
transverse resistivity $\rho_{xy}$ for the single crystal and thin
film of SrRuO$_3$, as well as for the Ca doped
Sr$_{0.8}$Ca$_{0.2}$RuO$_3$ thin film. The corresponding transverse
conductivity $\sigma_{xy}$ is shown in (D) as a function of the
magnetization, together with the results of first principle
calculations for cubic and orthorhombic
structures~\cite{orth-struc}. In our calculations, the change of
magnetization is taken into account by the rigid splitting of up and
down spin bands. As the transverse conductivity should vanish with $M$
at high temperatures, the calculated $\sigma_{xy}$ is multiplied by
the additional $M/M_0$ ($M_0$ = 1.5$\mu_B$) factor, which does not
affect its behavior except at the very vicinity of Tc.}
\end{figure}

\newpage
\begin{figure}
\includegraphics[scale=0.8]{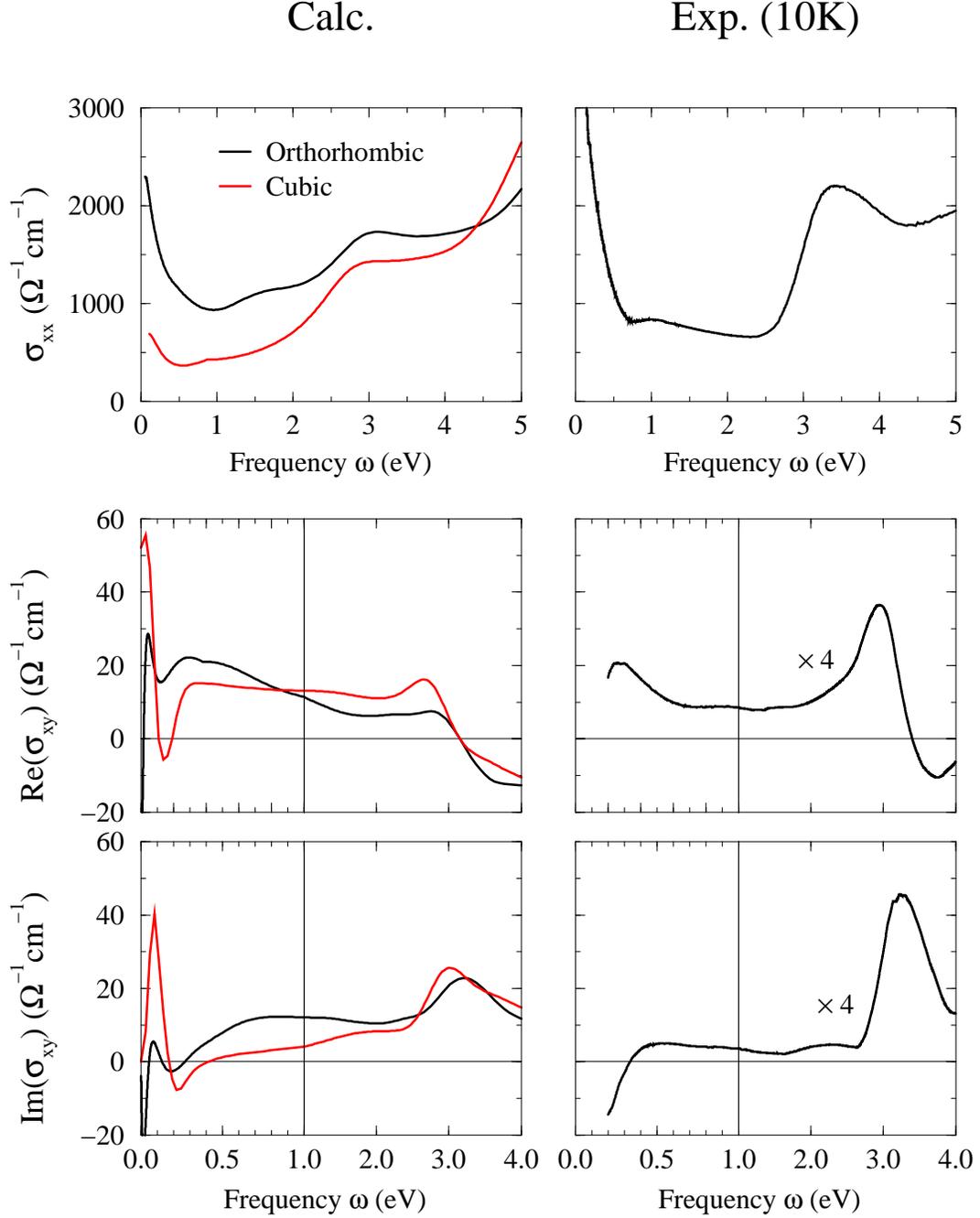}
\caption{Calculated (left panels) and measured (right
panels) longitudinal ($\sigma_{xx}$) and transverse ($\sigma_{xy}$)
optical conductivity of SrRuO$_3$ film. The measurements were
performed at low temperature (10K). The calculations were done for
both the orthorhombic single crystal structure and the hypothetic
cubic structure by keeping the average Ru-O bond length. The
experimental $\sigma_{xy}$ are shown by multiplying a factor of 4. The
quantitative comparison between the experiments and calculations about
the absolute value of $\sigma_{xy}$ would require more accurate
structure information (see text part). Nevertheless, the clear peak
structures for the low energy $\sigma_{xy}$ is the demonstration of
monopoles associated with $b_z({\bf k})$.}
\end{figure}

\newpage
\begin{figure}
\includegraphics[scale=0.7]{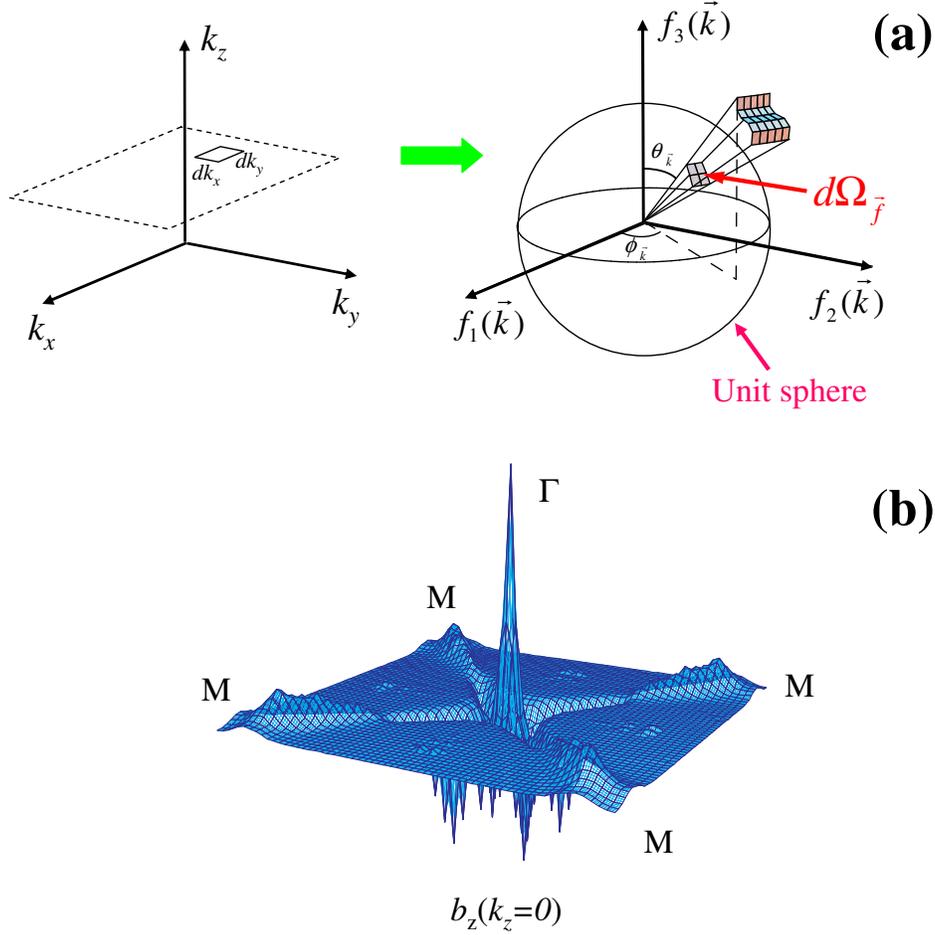}
\caption{(A) Geometrical meaning of the contribution
$\sigma_{xy}$ when the two bands are nearly degenerate. The
two-dimensional Hamiltonian matrix $H(\vec k)$ can be generally
written as $H(\vec k) = \sum_{\mu =0, 1,2,3} f_\mu(\vec k) \sigma_\mu
$ where $\sigma_{1,2,3}$ are the Pauli matrices and $\sigma_0 $ is the
unit matrix.  By mapping from $\vec k$ to the vector $\vec f(\vec k) =
( f_1(\vec k), f_2(\vec k), f_3(\vec k))= f({\vec k}) ( \cos
\varphi_{\vec f} \sin \theta_{\vec f}, \sin \varphi_{\vec f} \sin
\theta_{\vec f}, \cos \theta_{\vec f} ) $, the contribution to
$\sigma_{xy}$ from the neighborhood of degeneracy region can be given
by the ${\vec f}$-space solid angle $d \Omega_{\vec f}= [\partial
(\theta_{\vec f}, \varphi_{\vec f}) / \partial (k_x, k_y)] \sin
\theta_{\vec f} d k_x d k_y = d \varphi_{\vec f} \sin \theta_{\vec f}
d \theta_{\vec f}$ for the infinitesimal $d k_x d k_y$ integrated over
{\bf k}. The solid angle corresponds to the flux from the monopole at
${\vec f}={\vec 0}$. (See SOM). (B) Calculated flux distribution in
$\vec k$-space for $t_{2g}$ bands as a function of $(k_x,k_y)$ with
$k_z$ being fixed at 0 for SrRuO$_3$ with Cubic structure. The sharp
peak around $k_x=k_y =0$ and the ridges along $k_x = \pm k_y$ are due
to the near degeneracy of $d_{yz}$ and $d_{zx}$ bands because of the
symmetry reasons as explained in the online supporting text.}
\end{figure}

\newpage
\begin{figure}
\includegraphics[scale=1.2]{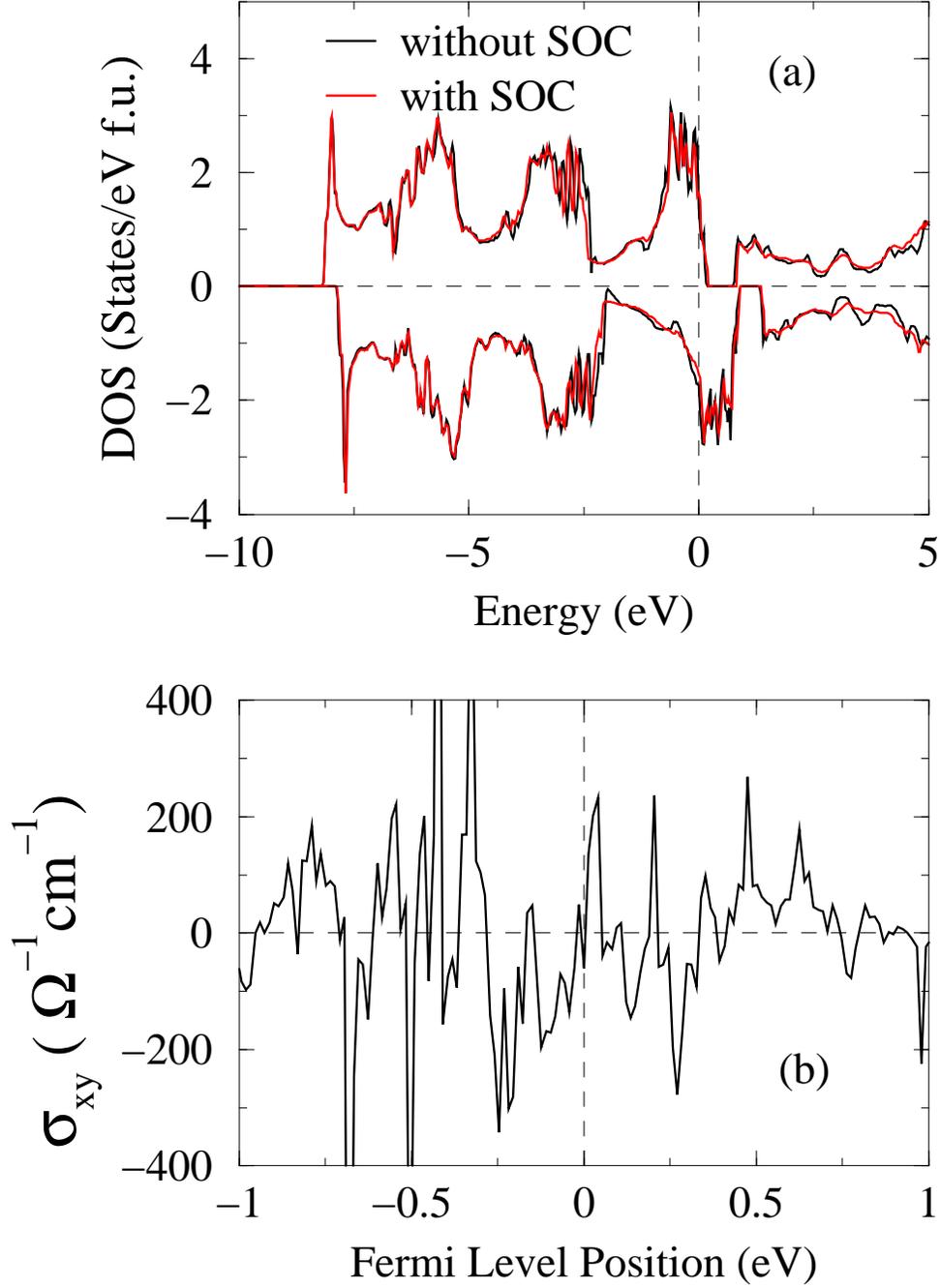}
\caption{The calculated (A) density of states (DOS) and
(B) $\sigma_{xy}$ as functions of Fermi level position for the
orthorhombic structure of single crystal SrRuO$_3$. The Fermi level is
shifted rigidly relative to the converged solution, which is specified
as zero point here. The sharp and spiky structure of $\sigma_{xy}$ is
the demonstration of the singular behavior of magnetic monopoles.}
\end{figure}

\clearpage
\newpage
\section{Supporting Online Material}

\section*{Materials and Experimental Methods}

Both stoichiometric SrRuO$_3$ and Ca-doped
(Sr$_{0.8}$Ca$_{0.2}$RuO$_3$) thin (500\AA) single crystal epitaxial
films were grown on the (001) surface of SrTiO$_3$ (STO) single
crystal substrate by pulsed laser deposition (PLD) employing KrF laser
pulses (100mJ) focused on the polycrystalline target, while SrRuO$_3$
bulk single crystal were prepared using a flux method (SrCl$_2$
flux). The quality of those samples were confirmed by the X-ray
diffraction. The bulk single crystal is orthorhombic, with
$a$=3.911\AA, $b$=3.936\AA, $c$=3.922\AA. The films are coherently
strained by the SrTiO$_3$ substrate, yielding a tetragonal distortion
in the [001] direction. As the results, the out-of-plane lattice
constants of films are elongated ($c$=3.950\AA\ for SrRuO$_3$ film and
$c$=3.931\AA\ for Ca-doped film), and the magnetic easy axis is
perpendicular to the film plane.

For the transport measurements of films, they were patterned in Hall
bar geometry using conventional photo-lithography and Ar ion
dry-etching. The Hall resistivity $\rho_{H}$ was measured together
with the longitudinal resistivity $\rho_{xx}$ as a function of
temperature under applied magnetic field. The anomalous resistivity
$\rho_{xy} $ was determined after subtraction of the ordinary Hall
contribution from the measured $\rho_{H}$, and the transverse
conductivity $\sigma_{xy}$ was estimated as $-\rho_{xy}/\rho^2_{xx}$.
In addition to the transport measurement, the frequency ($\omega$)
dependent conductivities were measured for SrRuO$_3$ film by optical
method. The longitudinal part of the optical conductivity
$\sigma_{xx}(\omega)$ was obtained from Kramers-Kronig transformation
of normal reflectivity from 0.08eV to 40eV, while the transverse part
$\sigma_{xy}(\omega)$ was deduced from magneto-optical Kerr
spectra. The spectra were measured by a polarization modulation
method.  For the energy region of 0.7 -- 4 eV, ordinary measurement
system with grating monochromator and photo-elastic modulator (CaF$_2$
window, modulation frequency of 56 kHz) was used for the measurement.
For the energy region below 0.8 eV, Fourier transform infrared (FT-IR)
spectrometer with rapid scan mode (scan speed 0.16 cm/s) was utilized
for the measurement.  The light from the FT-IR interferometer was, at
first, polarized by wire-grid polarizer (BaF$_2$ substrate).  Then the
polarization was modulated by ZnSe photo-elastic modulator (modulation
frequency of 50 kHz) and focused by BaF$_2$ lens (focal length of 150
mm).  The light reflected from the sample was detected by photovoltaic
type HgCdTe detector.  By this system we can measure magneto-optical
spectra down to 0.2 eV with accuracy higher than 0.01 degree.

\section*{Some Details of First-Principles Calculations} 

First-principles calculations of $\sigma_{xy}$ are quite challenging,
requiring the combination of several modern calculation techniques and
powerful supercomputer sources. The plane-wave pseudopotential
calculations~\cite{Rev} were performed based on the local spin density
approximation (LSDA), that give good descriptions for the electronic
and magnetic properties of this compound \cite{Mazin,Singh}. The
spin-orbital coupling (SOC) is treated self-consistently by using the
relativistic fully separable pseudopotentials~\cite{SOC} in the
framework of non-collinear magnetism formalism.  The inter-band
optical transitions are calculated from the converged Kohn-Sham eigen
states by using the Kubo formula~\cite{Wang,Optical}. The analytical
tetrahedron method~\cite{tetra} has been used for the accurate ${\vec
k}$-space integration, and the convergence has been checked
carefully. The finite life-time broadening $\delta$ has been estimated
from the experimental residual resistivity and extended Drude analysis
of $\sigma_{xx}$. The validity of our calculation techniques has been
well demonstrated in the explanation of anisotropic optical data in
Ca$_2$RuO$_4$~\cite{Fang2}.

\section*{$\sigma_{xy}$ and Magnetic Monopole in Momentum Space }

Here we present more details on the contribution to the transverse
conductivity $\sigma_{xy}$ from the ${\bf k}$ region where two bands
are nearly degenerate.  Enhancement of the gauge field $b_z(\vec k)$
occurs when more than two bands are close energetically, which
corresponds to the magnetic monopole and shows its fingerprint in
anomalous Hall effect (AHE) and Kerr rotation as described
below~\cite{onoda2}.  Consider the general case where two band
Hamiltonian matrix $H(\vec k)$ at $\vec k$ can be written as $H(\vec
k) = \sum_{\mu =0, 1,2,3} f_\mu(\vec k) \sigma_\mu $ where
$\sigma_{1,2,3}$ are the Pauli matrices and $\sigma_0 $ is the unit
matrix.  Then we can consider the mapping from $\vec k$ to the vector
$\vec f(\vec k) = ( f_1(\vec k), f_2(\vec k), f_3(\vec k)) = f({\vec
k}) ( \cos \varphi_{\vec f} \sin \theta_{\vec f}, \sin \varphi_{\vec
f} \sin \theta_{\vec f}, \cos \theta_{\vec f} )$ as shown in Fig. 3A
of the main text.  Then $H(\vec k)$ can be easily diagonalized to obtain
the two eigenvalues $\varepsilon_{\pm}(\vec k) = f_0(\vec k) \pm
f(\vec k)$. Calculating Eq.5 of the main text in this case, we obtain the
contribution to $\sigma_{xy}$ from these two bands as
\begin{eqnarray*}
\ \ \ \ \ \ \ \ \ \ \ \ \ \ \ \ \ \ \ \ \ \ \
\sigma_{xy}^{\rm{2-bands}} &=&
{ {e^2} \over { 8 \pi h} } \int d^3 {\vec k} 
[ n_F( \varepsilon_-(\vec k)) - n_F (\varepsilon_+ (\vec k) ) ]
\nonumber \\
&\times&
\biggl( \frac{ \partial \varphi_{\vec f} } {\partial k_{x}} 
\frac{\partial \theta_{\vec f}}{\partial k_{y}} 
-\frac{ \partial \varphi_{\vec f} }{\partial k_{y} } 
\frac{ \partial \theta_{\vec f} } {\partial k_{x} } \biggr) 
{\rm sin}\theta_{\vec f} 
\nonumber \\
&=& 
{ {e^2} \over { 8 \pi h} } \int d k_z d \Omega_{\vec f} 
[ n_F( \varepsilon_-(\vec k) ) - n_F (\varepsilon_+ (\vec k) )]. 
\ \ \ \ \ \ \ \ \ \ \ \ \ \ \ \ \ \ \ \ \ \ \ \ \ 
{\rm (S1)}
\nonumber
\end{eqnarray*}
Here  $d \Omega_{\vec f}= [\partial (\theta_{\vec f}, \varphi_{\vec
f}) / \partial (k_x, k_y)] \sin \theta_{\vec f} d k_x d k_y
= d \varphi_{\vec f} \sin \theta_{\vec f} d \theta_{\vec f} $ 
is the $\vec f$-space solid angle, which is the integral of the gauge field 
\begin{eqnarray*}
\ \ \ \ \ \ \ \ \ \ \ \ \ \ \ \ \ \ \ \ \ \ \ \ \ \ \ \ \ \ \ \ 
\ \ \ \ \ \ \ \ \ \ \ \ \ \ \ \ \ \ \ \ \ \ \ \ 
{\vec b} (\vec f) = \pm { {\vec f } \over 
{ | {\vec f} |^3} }
\ \ \ \ \ \ \ \ \ \ \ \ \ \ \ \ \ \ \ \ \ \ \ \ \ \ \ \ \ \ \ 
\ \ \ \ \ \ \ \ \ \ \ \ \ \ \ \ \ \ \ \ \ \ \ \ \ \ \ \ \ \ \ 
{\rm (S2)}
\end{eqnarray*}
due to the monopole at ${\vec f} = {\vec 0}$ over the infinitesimal
surface in $\vec f$-space corresponding to the small square $d k_x d
k_y $ in $\vec k$-space (Fig.3A of the main text).  Therefore
$\sigma_{xy}^{\rm{2-bands}}$ again has the geometrical meaning in
$\vec f$ space.  This gauge field strongly depends on $\vec k$ in the
(near) degeneracy case, i.e., when ${\vec f}(\vec k)$ is near the
monopole.

There are two cases for the (near) degenerate bands. One is the
accidental degeneracy \cite{wigner,herring} where the three equations
$f_1(\vec k)= f_2(\vec k) = f_3(\vec k) =0$ are satisfied at ${\vec k}
= {\vec k}_0$.  Near this accidental band crossing, one can expand as
$f_a(\vec k) = \sum_{b} \alpha_{a b} ( k_b - k_{0 b})$ where $a,b =
1,2,3$.  In this case the $\vec f$-space can be identified with the
$\vec k$-space, and the gauge field distribution in $\vec k$-space
around ${\vec k} = {\vec k}_0$ is similar to Eq.S1 above replacing
$\vec f $ by ${\vec k} - {\vec k}_0$.  Although there occurs no
singularity in $\sigma_{xy}$ at $\mu = \varepsilon_\pm ({\vec k}_0)$
due to the cancellation between the positive and negative $k_z - k_{z
0}$, the nonlinear dependence of $f_a({\vec k})$ gives rise to the
strong $\mu$-dependence of $\sigma_{xy}$ slightly away from this
energy.  The other class of (near) degeneracy is due to the symmetry,
where the ${\vec k}$-group has the irreducible representation with the
dimensions more than 2. As an example, one can consider the simplest
tight binding model of $t_{2g}$ orbitals $d_{yz}$ and $d_{zx}$ on the
cubic perovskite structure, which is relevant to the SrRuO$_3$
discussed in the main text.  In this case the $H(\vec k)$ for these two bands
with up-spin is given by $f_0(\vec k) = - 2 t_1 \cos k_z - t_1 ( \cos
k_x + \cos k_y )$, $f_1(\vec k) = 2 t_2 \sin k_x \sin k_y$, $f_2(\vec
k) = - \lambda M $, $f_3(\vec k) = - t_1 ( \cos k_x - \cos k_y )$,
where $t_1, t_2$ are the effective intra- and inter-orbital transfer
integrals respectively, $\lambda$ is the spin-orbit coupling (SOC)
constant, and $M$ is the magnetization. When $\lambda M = 0$, there
occurs the degeneracy along the line $\vec k = (0,0,k_z)$. Furthermore
when $t_2=0$, there occurs the degeneracy along the plane $k_x = \pm
k_y$.  Considering the case $t_1 >> t_2 >> \lambda M$, which is
relevant to SrRuO$_3$, the gauge field $b_z(\vec k)$ with $k_z=0$ has
a largest peak at $k_x=k_y=0$ and is enhanced along the lines $k_x =
\pm k_y$, which is actually seen in the realistic calculation for
SrRuO$_3$ shown in Fig.3B of the main text.  In this case there occurs no
cancellation of $b_{n z}(\vec k)$ from the integral over $k_z$.
 
Although the discussion above is applicable to any (near) degeneracy,
the (singular) gauge field from different band crossings in ${\vec
f}$-space cancel out in the presence of the time-reversal symmetry
and/or the absence of the SOC. The former prohibits the finite
$\sigma_{xy}$.  In the absence of SOC, up-spin and down-spin bands are
decoupled, and each of them can be represented by the Hamiltonian
matrix without the time-reversal symmetry breaking.  In ferromagnets
with the SOC, the singular behavior of $\sigma_{xy}$ by changing some
parameters such as the chemical potential and the magnetization is the
fingerprint of the monopoles in $\vec k$- and/or $\vec f$-space.


\begin{thebibliography}{}
\bibitem{berry}It has been recognized in the original paper by
M. V. Berry (Proc. R. Lond. {\bf{A392}}, 45 (1984)) that the
degeneracy point in the parameter space acts as a magnetic monopole
where the gauge field is enhanced.
\bibitem{real} Note that this gauge field is distinct from that
of the magnetic field ${\vec B}({\vec r})$ in real space, 
although they are analogous to each other. 
In the presence of ${\vec B}({\vec r})$, the covariant momentum operator
${\pi_\mu}$ is given by $\pi_\mu = - i \partial_{x_\mu} + e A_{\mu}({\vec r})$
where ${\vec B} = \nabla \times {\vec A}$. This leads to the
commutation relation 
$[\pi_x, \pi_y] = - ie ( \partial_x A_y - \partial_y A_x ) = - ie B_z$
etc., and to the Lorenz force due to the magnetic field ${\vec B}$.   
Therefore these two gauge fields, i.e., $b_\mu({\vec k})$ and
${\vec B}({\vec r})$, are dual to each other, and 
the presence of the one does not necessarily mean that of the other.
\bibitem{Ca-more}More Ca-doped samples with different concentrations
have been measured. They all follow the same trend, and are not shown
here due to the space.
\end{thebibliography}
\end{document}